\documentclass[a4paper,10pt]{article}
\usepackage[utf8]{inputenc}
\usepackage{authblk}
\usepackage{cite} 
\usepackage[mathscr]{eucal}
\usepackage{amsfonts}
\usepackage{physics}

\begin{document}
\title{Weak gravitational quantum effects in boson particles}
\author[1]{Omar Gallegos\thanks{ogallegos@fis.cinvestav.mx}}
\author[1]{Tonatiuh Matos\thanks{tmatos@fis.cinvestav.mx}}
\affil[1]{\small{\textit{Departamento de Física, Centro de Investigación y de Estudios Avanzados del IPN, A.P. 14-740, 07000 CDMX, México.}}}

\maketitle
\begin{abstract}
We rewrite the Klein-Gordon (KG) equation in an arbitrary space-time transforming it into a generalized Schr\"odinger equation. Then, we take the weak field limit and show that this equation has certain differences with the traditional Schr\"odinger equation plus a gravitational field. Thus, this procedure shows that the Schr\"odinger equation derived in a covariant manner is different from the traditional one. We study the KG equation in a Newtonian space-time to describe the behavior of a scalar particle in an inertial system. This particle is immersed in a gravitational field with the new Schr\"odinger equation. We study particular physical systems given examples for which we find their energy levels, effective potential and the wave function of the systems. The results contain the gravitational effects due to the curvature of space-time. Finally, we discuss the possibility of the experimental verification of these effects in a laboratory using non-inertial reference frames. 
\end{abstract}

\section{Introduction}

\qquad In the last century, General Relativity (GR) and Quantum Mechanics (QM), the two pillars of modern physics, have been developed and verified independently with great precision, while quantum physics describes successfully the behavior of tiny particles, GR is very accurate for forces at cosmic scales. However, in some cases, the two theories produce incompatible results which give rise to different definitions for the same concept. We think the inconsistency between GR and QM owes to the concept of interaction between particles. In QM two particles interact when they exchange a virtual particle, while in GR the interaction is just due to the geometry of space-time. In this work, we adopt the geometrical GR concept, instead of the exchange of virtual particles, to test if a boson gas follows the Klein- Gordon equation in a curved space-time\cite{Birrell:1982}, and to test GR in the quantum regime. These results could be useful either for laboratory particles as well as for the study of the quantum character of boson particles proposed as dark matter (see for example \cite{Matos1:1998vk}, \cite{Hui:2016ltb}).\newline 

An important problem in theoretical and fundamental physics is not to have a Theory of Everything where the principal theories in physics, GR and QM, can be compatible. In the last decades, some theories have been proposed \cite{Loop},\cite{String} to this unification. Nevertheless, the experimental verification of these candidates and their theoretical problems are so far too complex. Several physicists have tried to test if gravity has a quantum nature with different proposals for experiments and observations\cite{Casimir}\cite{entaglement}, but some proposals are not feasible with the technology available to this day. We do not pretend to propose a new Theory of Everything, instead, we want to give a new different way to measure the gravitational effect due to the curvature of space-time on quantum systems, especially on scalar particles. These results do not definitely prove if gravity is or not a quantum interaction but they present a closer path to answer this fundamental question.\newline

Einstein Equivalence Principle (EEP), one of the foremost ideas for developing General Relativity, where the concept of inertia takes a different role from the one used in Newtonian mechanics. EEP states that within a closed laboratory freely falling in a uniform gravitational field, experiments would be indistinguishable from a laboratory in a state of uniformly accelerated motion. All these dynamical experiments yield the same results as obtained in an inertial state of uniform motion unaffected by gravity.\cite{Will_2014}\cite{Wald:1984}\cite{Relativity:2009} Hence, we use this principle to develop this work, studying some different examples of QM on an inertial frame immersed in a gravitational field. To measure these effects in a laboratory, we will place a quantum system on a non-inertial frame, hoping to get the same results both in the theoretical part and in the experimental one. Furthermore, we expect to obtain results on quantization as in QM. Thus, one of the objectives of this paper is to test the EEP in quantum scales. Thus, a proof of GR on a quantum scale.\newline

We would expect that there exists a regime where gravity aspects on a quantum system are detectable. With a simple dimensional analysis we may have an idea where the gravitational effects on a quantum system are important, i.e. the physical scale where these effects can be observed. It is possible to compare two well-known quantities in QM and GR, which are the Bohr radius $r_{\mathrm{bohr}}$ and the Schwarzschild radius $r_{\mathrm{sch}}$. Both radii are defined as $r_{\mathrm{bohr}}=\hbar/(\mu c \alpha)$ and $r_{\mathrm{sch}}=2GM/c^2$ where $c$, $G$, $\alpha$ are the speed of light, the gravitational constant and the fine-structure constant, respectively. Also $\hbar=h/2\pi$ is the Planck constant, and $M$ is the mass that produces the gravitational field, while $\mu$ is the reduced mass of a couple of particles. In a case where the comparison between masses is such that $M>>m$, it leads us to the approximation $\mu\approx m$, where $m$ is the mass of a boson particle, which is inside a gravitational field generated by a source of mass $M$. Thereby, the following expression is obtained when both radii $r_{\mathrm{bohr}}\sim r_{\mathrm{sch}}$ are comparable 
\begin{equation}
\label{dimensional analysis}
M\sim\dfrac{\hbar c}{2G\alpha m}=\dfrac{m_{pl}^2}{2\alpha m}\approx\dfrac{3.25\times10^{-14}}{m}\text{kg}^2,
\end{equation}
where $m_{\mathrm{pl}}=\sqrt{\hbar c/G}$ is the Planck mass. This result was inspired by the micro black holes described in\cite{microBH} and it enables us to find the limit for measuring quantum gravitational effects. If we assume $m=m_e$ is the mass of an electron, we obtain that $M\sim 3.57\times10^{16}$kg is the mass for which a particle with the electron mass should feel a gravitational effect. On the other hand, the mass of Earth is $\sim 5\times 10^{24}$kg, namely, we should measure in Earth the quantum gravitational effects on an electron. Although, we consider an electron as a test particle, due to it is a standard mass measurement. If it can feel these gravitational quantum effects with this dimensional analysis, we also think other heavier particles feel them, like alpha particles whose mass is $6.64\times 10^{-27}$kg. Thus, for these particles $M\sim 4.9\times 10^{12}$kg. However, experimentally, to measure only the gravitational effects on a quantum system, and not other interactions as electromagnetic, it would be better to use neutral or scalar particles, as they can be either neutron or alpha systems. Now, if we consider the mass of an ultralight scalar particle $\sim 10^{-22}\mathrm{eV/c^2}$, the mass proposed in models of Scalar Field Dark Matter (SFDM)\cite{Matos1:1998vk}, \cite{Hui:2016ltb}, we calculate from Eq.(\ref{dimensional analysis}) that $M\sim 10^{12}M_\odot$ (the galaxies have a mass of this order). Thus, we can say, if dark matter is a scalar field with mass $m\sim 10^{-22}\mathrm{eV/c^2}$ \cite{Matos1:1998vk}, \cite{Hui:2016ltb}, the galaxy should present a quantum behavior because here, SFDM obeys the KG equation. Therefore, we study well-known examples of QM for scalar particles with corrections due to the curvature of space-time.\newline

We start from the KG equation in curved space-time for finding a method to compare the eigenvalues with and without space-time curvature. However, there exist previous works about the solutions for the KG equation in Schwarzschild background space-time\cite{Lehn:2018},\cite{KGinSch:1976}. Authors on \cite{Lehn:2018} give a numerical analysis for a free mass particle into the KG equation in a Schwarzschild background space, then they add the approximation for the standard hydrogenic modified radial wave functions.\newline

On the other hand, authors on \cite{KGinSch:1976} find an exact solution for the same problem on \cite{Lehn:2018} by the Liouville-Green method. If we want to add different potentials, we need to use an approximation method but this proposed solution, here, is only valid for a free particle in a Schwarzschild background space, which has the like-hydrogen atom problem.\newline 

Nevertheless, to consider distinct potentials, as usually in QM problems, it would be needed to do a similar work of the authors on \cite{Lehn:2018},\cite{KGinSch:1976} for each different problem. While, we consider the gravitational effects as perturbations of these systems. Furthermore, we use the perturbation theory as in  QM to solve our generalized Schr\"odinger equation that is deduced in the next chapter. Thus, we can give a comparison way to the solutions into a flat and curved space-time with an external potential, where in some cases the external potential will be the perturbation of the gravitational solution and in others taking gravitational effect as a perturbation of the system with an external potential.

\section{Generalized Schr\"odinger Equation}

\qquad In order to analyze the gravitational effect due to curvature of space-time in a quantum system, we focus on a scalar field following reference \cite{Matos:2016uxo}, in which the KG covariant equation with an external potential is described 
\begin{equation}
\label{KG covariant}
\Box\Phi-\dfrac{\mathrm{d}\mathscr{V}}{\mathrm{d}\Phi^*}=0,
\end{equation}
here $\Box=g^{\mu\nu}\nabla_\mu\nabla_\nu=\dfrac{1}{\sqrt{-g}}\partial_\mu(\sqrt{-g}g^{\mu\nu}\partial_\nu)$ is the D'Alembertian operator associated to an arbitrary metric $g_{\mu\nu}$, now $\Phi=\Phi(t,\vec{x})$ is the scalar field,  $\Phi^*=\Phi^*(t,\vec{x})$ is its conjugated complex and the scalar field potential is $\mathscr{V}=\mathscr{V}(\Phi,\Phi^*)$ endowed with an external potential $V$ just as it is shown in the following equation
\begin{equation}
\label{potential}
    \mathscr{V}=\left(\dfrac{m^2c^2}{\hbar^2}+\dfrac{\lambda n_0}{2}+\dfrac{2m}{\hbar^2}V\right)\Phi\Phi^*,
\end{equation}
where $m$ is the mass of the scalar field and $n_0=n_0(t,\vec{x})$ is defined as the scalar field density, namely $n_0=|\Phi|^2$. The factor in front of the third term is due to dimensional analysis, with that we return to the standard Schrödinger equation, as we see below. With this analysis we could find some other potential forms of higher order, nevertheless we did not take into account these expressions because their contributions are smaller than the linear term. On the other hand, the space-time is expanded in 3+1 slices, such that the coordinate $t$ here is the parameter of evolution, the 3+1 metric then reads
\begin{equation}
\label{metric}
	\mathrm{d}s^2 = - N^2 c^2 
	\mathrm{d}t^2 + \gamma_{ij} \left(\mathrm{d}x^i + N^i c \, \mathrm{d}t \right) \left(
	\mathrm{d}x^j + N^j c \, \mathrm{d}t \right) ,
\end{equation}
$N$ represents the lapse function which measures the proper time of the observers traveling along the world line, $N^i$ is the shift vector that measures the displacement of the observers between the spatial slices and $\gamma_{ij}$ is the spatial metric.\newline

The KG equation is a covariant equation whose origin lies in quantum field theory. We can obtain a  Schr\"odinger equation, starting from equation (\ref{KG covariant}) with the potential in eq.(\ref{potential}), using the metric in eq.(\ref{metric}) and transforming the scalar field by $\Phi(t,\vec{x})=\Psi(t,\vec{x}) e^{-i\omega t}$. Following \cite{Matos:2016uxo} we obtain
\begin{eqnarray}
	i \nabla^0 \Psi - \frac{1}{2\omega} \Box_G \Psi 
    + \frac{1}{2\omega} \left( \widetilde{m}^2 + \lambda n_0+\dfrac{2m}{\hbar^2}V \right) \Psi &&
     \nonumber\\
	+\quad \frac{1}{2} \left( - \frac{\omega}{N^2}
	+ i \, \Box_G \, t \right) \Psi &=& 0 ,
\label{eq:GP}
\end{eqnarray}
being $\lambda$ the coupling parameter. Here $\omega=mc^2/\hbar$ is the characteristic frequency of the scalar field. The D'Alembertian operator is $\Box_G=\nabla^\mu\nabla_\mu$, which are associated to the metric in eq.(\ref{metric}), and $\widetilde{m}$ stands for the mass in units where $c=\hbar=1$, i.e, $\widetilde{m}=m^2c^2/\hbar^
2$ (see \cite{Matos:2016uxo}). We can interpret the function $\Psi$ as a wave function analogously as in QM.\newline

This generalized Schrödinger equation (SE) is relevant because we can study it in an arbitrary frame, since it was obtained in a covariant manner. In some limits we can return to the traditional SE as we will see below.\newline 

Beginning with eq.(\ref{KG covariant}) but now using the $\Psi$ variable, we obtain Eq.(\ref{eq:GP}) using the 3+1 metric (\ref{metric}) and the potential (\ref{potential}). Equation (\ref{eq:GP}) can be interpreted as the covariant generalization of the Schr\"odinger equation for any curved space-time, (see\cite{Chavanis:2016shp} and \cite{Matos:2016uxo}) where this equation in the weak field limit reduces to the standard Schr\"odinger one.\newline

Hereafter, we use the Newtonian geometry given by the Newtonian metric 
\begin{equation}
\label{Newtonian metric}
\mathrm{d}s^2=-\left(1-\dfrac{2GM}{rc^2}\right)c^2\mathrm{d}t^2+\left(1+\dfrac{2GM}{rc^2}\right)\mathrm{d}x_i\mathrm{d}x^i,
\end{equation}
Newtonian gravity is known to be valid when the gravitational fields are weak, that is $GM/rc^2<<1$. Using the Newtonian metric in Eq.(\ref{eq:GP}) yields
\begin{align}
\label{KG-N equation}
 \dfrac{\hbar^2}{2m}\Box_N\Psi&+\left(1+\dfrac{2U}{c^2}\right)^{-1}\left(i\hbar\dfrac{\partial \Psi}{\partial t}+\dfrac{mc^2}{2}\Psi\right)\nonumber\\
  &-\dfrac{\hbar^2}{2m}\left(\widetilde{m}+n_0\lambda+\dfrac{2m}{\hbar^2}V\right)\Psi=0,  
\end{align}
where $U=-GM/r$ is the gravitational potential, hither the D'Alembertian operator $\Box_N$ is related to the metric given by eq.(\ref{Newtonian metric}). Furthermore, we do not consider self-interaction, here the contribution of the term $\lambda$ is negligible.\newline 

Additionally, terms of equal or greater order than $\left(\dfrac{2U}{c^2}\right)^2$ can be ignored. With all this in mind, the linearization of the generalized Schr\"odinger equation (\ref{KG-N equation}) gives rise to the following equation
\begin{equation}
\label{KG-N motion equation}
-\dfrac{\hbar^2}{2m}\nabla^2\Psi+V\Psi+mU\Psi +\left(\dfrac{2U}{c^2}V-\dfrac{2\hbar^2U}{mc^2}\nabla^2\right)\Psi=i\hbar\dfrac{\partial\Psi}{\partial t}-\dfrac{\hbar^2}{2mc^2}\dfrac{\partial^2\Psi}{\partial t^2}.
\end{equation}

Henceforth, we are going to work with the Laplacian operator $\nabla^2=\nabla\cdot\nabla$ in flat space, specially for the examples in the following chapters where we use spherical symmetry. In other words, equation (\ref{KG-N motion equation}) comes from the KG equation in a curved space-time immersed in a weak gravitational field, which is described by a Newtonian geometry.\newline

Focusing on the comparison of the traditional Schr\"odinger equation plus a gravitational potential, we note that now there are two additional terms inside the parenthesis in eq.(\ref{KG-N motion equation}) for the spatial part. Furthermore, the time-dependence part in eq.(\ref{KG-N motion equation}) has an extra second derivative term in comparison to the standard Schr\"odinger equation in QM.  These terms were not simply added in the Schr\"odinger equation, they appear from the covariant meaning of the equation. From a qualitative point of view, this means that the QM version of the interactions between particles fulfills the Schr\"odinger equation, while the GR version of these interactions fulfills the generalized Schr\"odinger eq.(\ref{KG-N motion equation}). This difference is the main idea of this work. In what follows, we calculate the quantum quantities for different external potentials of well-known problems in QM and compare them with those corresponding to GR. For the subsequent chapters our equation of motion will be given by eq.(\ref{KG-N motion equation}).\newline

We can solve the time-dependence of the equation (\ref{KG-N motion equation}) using the separation of variables method, we can express the wave function as $\Psi=\Psi(x^\mu)=\Psi(t)\Psi(x^i)$, for $\mu=0,1,2,3$ and $i=1,2,3$. Thus, we need to solve the following differential equation 

\begin{equation}
\label{time-dependece diff eq}
    i\hbar\dfrac{\partial\Psi(t)}{\partial t}-\dfrac{\hbar^2}{2mc^2}\dfrac{\partial^2\Psi(t)}{\partial t^2}=E\Psi(t).
\end{equation}

It is easy obtain a general solution to this differential equation, which can be expressed as follows
\begin{equation}
    \label{time solution}
    \Psi(t)=A\exp(i\omega_+t)+B\exp(i\omega_-t),
\end{equation}
where $A$ and $B$ are constants. Moreover $\hbar\omega_+=mc^2\left(1-\sqrt{1+2E/(mc^2)}\right)$ and $\hbar\omega_-=mc^2\left(1+\sqrt{1+2E/(mc^2)}\right)$. Since, we want to return to the time-dependence solution of the standard Schr\"odinger equation in certain limit. We use the Taylor expansion when $E<<mc^2$ to find $A$ and $B$. The wave-function $\Psi(x^\mu)$ must be normalized. Therefore, if we want to return to the traditional Schr\"odinger equation, as just as, in QM, we impose $A=0$ and $B=1$. Note that the difference in time-dependence part in our generalized Schr\"odinger equation does not affect the eigenvalues of the generalized Schr\"odinger equation, it only contributes to the time part of wave function as $\Psi(t)=\exp(i\omega_-t)$, instead of $\exp(-iEt/\hbar)$. Nevertheless, we get back to QM case taking the previously mentioned limit. The modification of the eigenvalues is due to the extra terms in the spatial part of the Schr\"odinger equation. \newline 

With the goal of testing EEP in a quantum region, we are going to show in the subsequent chapters distinct calculations about well-known examples in QM. In them, the correction terms due to the curvature of space-time, which were introduced in the KG covariant equation derived in eq.(\ref{KG-N motion equation}), are going to appear. To compare the cases with and without a gravitational source, the extra terms in eq.(\ref{KG-N motion equation}) will be taken as perturbations in the Schr\"odinger equation, for this reason we decide to opt for the QM formalism, using perturbation theory to give a clearer comparison between the cases with and without gravitational field. The results of the gravitational field contribution have been well studied in the standard literature\cite{Cohen:2005}\cite{Cohenv2:2005}\cite{Zettilli:2009}.

\section{Free Particle}

\qquad Firstly, we consider a free scalar particle with mass $m$ under the influence of a gravitational field generated by a source of mass $M$. Using an external potential $V_{\mathrm{free}}=0$ in the equation of motion (\ref{KG-N motion equation}), we have that
\begin{equation}
-\dfrac{\hbar^2}{2m}\nabla^2\Psi+mU\Psi-\dfrac{2\hbar^2U}{mc^2}\nabla^2\Psi=E\Psi,
\end{equation}
we recall that the gravitational potential is defined by $U=-GM/r$ and that an extra term $-\dfrac{2\hbar^2U}{mc^2}\nabla^2\Psi$ appears in the Schr\"odinger equation, although the treatment on these equations will be in the QM formalism. Using perturbation theory from the previous equation, we can write a principal Hamiltonian operator $\hat{H}_0$ and a perturbed Hamiltonian operator $\hat{H}_p$ in the following way 
\begin{align}
\label{H0 free particle}
&\hat{H}_0\Psi=-\dfrac{\hbar^2}{2m}\nabla^2\Psi+mU\Psi=-\dfrac{\hbar^2}{2m}\nabla^2\Psi-\dfrac{GMm}{r}\Psi,\\
\label{Hp free particle}
&\hat{H}_p\Psi=-\dfrac{2\hbar^2U}{mc^2}\nabla^2\Psi=\dfrac{4U}{c^2}(E_n^{(0)}-mU)\Psi.
\end{align}

In general, to find the corrections to eigenvalues of energy due to a perturbation in QM, the following expression is used
\begin{equation}
\label{energy correction}
E_n=E_n^{(0)}+\bra{\psi^{(0)}_n}\hat{H}_p\ket{\psi^{(0)}_n}+\sum_{m\neq n}\dfrac{|\bra{\psi^{(0)}_m}\hat{H}_p\ket{\psi^{(0)}_n}|^2}{E_n^{(0)}-E_m^{(0)}}+...,
\end{equation}
we can associate the eq.(\ref{energy correction}) as a series of contributions of higher order correction of energies $E_n=E_n^{(0)}+E_n^{(1)}+E_n^{(2)}+...$, where $E_n^{(j)}$ is the $j$th-order correction of the eigenvalues in the $n$th-state  of energy, hence we can say that the first-order correction is given by $E_n^{(1)}=\bra{\psi^{(0)}_n}\hat{H}_p\ket{\psi^{(0)}_n}$. The zero-order correction for eigenvalues of energy $E_n^{(0)}$ and the eigenfunctions are the well-known exact solutions of standard QM\cite{Cohen:2005,Cohenv2:2005,Zettilli:2009,Pena:2006}. Equation (\ref{energy correction}) gives the correction for a non-degenerated quantum system, though first-order correction is valid for both cases (degenerate and non degenerate systems). We do not present an expression for the second order correction in a degenerate system because the corrections we make are at most of first-order. Nevertheless, such a case can be found in standard QM textbooks.\\

Going back to our example of a free scalar particle inside a gravitational field, the first order correction of energy using the perturbed Hamiltonian from the hydrogen atom-like problem in QM is
\begin{align}
E_n&\approx E^{(0)}_n\left(1-\dfrac{4GM}{\rho_0c^2n^2}+...\right),
\\&=E^{(0)}_n\left(1+\dfrac{8E^{(0)}_n}{mc^2}+...\right)\notag,
\end{align}
being $E^{(0)}_n$ the well known energy for the hydrogen atom, that is 
\begin{equation}
\label{energy atom gravitation}
E^{(0)}_n=-\dfrac{(GM)^2m^3}{2\hbar^2n^2}=-\dfrac{GMm}{2\rho_0n^2},
\end{equation}
where $\rho_0=\hbar^2/(GMm^2)$ is the Bohr radius for the gravitational case. Thus, we expect that the gravitational field modifies the energy of a free particle in a quadratic level, suppressed by the rest energy of the boson particle. This result is surprising for an ultra-light dark matter boson \cite{Matos:1998vk}\cite{Matos:2000ss}, because this model postulates a boson particle with a mass of the order of $10^{-22}$eV$/c^2$. With such a mass the self-gravitation effects of the boson field are important in a system of particles, a feature that a heavy boson system does not present. 

\section{Isotropic Harmonic Oscillator}

\qquad Another important example, not only in QM but in physics as a whole, is the study of the harmonic oscillator potential. With this in mind, we analyze the case of potential $V_{\mathrm{osc}}$ for an isotropic harmonic oscillator. This analysis can be done in two ways. The first one is to start from eq.(\ref{KG-N motion equation}) with the principal potential given by the gravitational type $U=-GM/r$ and taking the isotropic harmonic oscillator $V_{\mathrm{osc}}=\dfrac{1}{2}m\omega_0^2r^2$ as perturbation. The other way is to regard the principal Hamiltonian as that of an isotropic harmonic oscillator, while the perturbation is taken from the gravitational potential. Therefore, equation (\ref{KG-N motion equation}) with the potential $V_{\mathrm{osc}}$, transforms into
\begin{equation}
-\dfrac{\hbar^2}{2m}\nabla^2\Psi+mU\Psi +\dfrac{1}{2}m\omega_0^2r^2\Psi+\left[\dfrac{2U}{c^2}\left(\dfrac{1}{2}m\omega_0^2r^2\right)-\dfrac{2\hbar^2U}{mc^2}\nabla^2\right]\Psi=E\Psi.
\end{equation}

\subsection{Harmonic Oscillator inside a Gravitational Field}

\qquad We start with the first cases exposed previously, we suppose that one has an isotropic harmonic oscillator immersed in a gravitational field, that means, $V_{\mathrm{osc}}<<U$. To solve this problem we can take the principal Hamiltonian operator $\hat{H}_0$ with the gravitational part and the perturbed Hamiltonian $\hat{H}_p$ in the following way
\begin{align}
\label{principal H IHO in curved space}
\hat{H}_0\Psi&=-\dfrac{\hbar^2}{2m}\nabla^2\Psi-\dfrac{GMm}{r}\Psi,\\
\label{Hp IHO in curved space}
\hat{H}_p&=\dfrac{1}{2}m\omega_0^2r^2-\dfrac{GMm\omega_0^2r}{c^2}-\dfrac{4GME^{(0)}_n}{rc^2}.
\end{align}

We are interested in the wave function with spherical symmetry $\Psi=\Psi(t,r,\theta,\phi)$. Thus, we apply the separation of variables method for $\Psi=$R$_{nl}(r)$Y$_{lj}(\theta,\phi)\Psi(t)$, where $\Psi(t)$ is the solution in eq.(\ref{time solution}), Y$_{lj}$ are the spherical harmonics and R$_{nl}(r)$ is the radial function for the hydrogen atom problem. Here $n,l$, and $j$ play the role of quantum numbers as in QM. The $\hat{H}_0$ Hamiltonian contains the well-known solutions of the eigenvalues from the hydrogen atom problem in terms of the recurrence relation for powers of $r$. It can be shown that the correction for first order of energy from eq.(\ref{energy correction}) is given by 
\begin{align}
E_n=&E^{(0)}_n\left[1-\dfrac{8E^{(0)}_n}{mc^2}-\dfrac{\omega_0^2\rho_0^2}{c^2}(3n^2-l(l+1))\right]
\\ &+\dfrac{1}{4}m\omega_0^2\rho_0^2\left[n^2(5n^2+1-3l(l+1))\right],\notag
\end{align}
where $E^{(0)}_n$ is given by eq.(\ref{energy atom gravitation}). Note that if $\omega_0=0$, we return to the case of a free particle on a gravitational field. Observe that as in the previous case, the modifications due to the gravitational field are of second order, but now, an additional term is added, which is proportional to the mass and the frequency $\omega_0$. In this case, the second term of the contributions of the perturbations of the gravitational potential due to the harmonic oscillator are negligible for ultra-light masses. Nevertheless, if it is a massive boson, then the quadratic contributions are not important. In any case, for any boson mass there is a contribution of the harmonic oscillator that must be taken into account.  

\subsection{Gravitational Field inside Harmonic Oscillator}

\qquad On the other hand, assuming the gravitational field as a perturbation in an isotropic harmonic oscillator with spherical symmetry, the principal Hamiltonian operator $\hat{H}_0$ is given by
\begin{align}
\hat{H}_0\Psi&=-\dfrac{\hbar^2}{2m}\nabla^2\Psi+\dfrac{1}{2}m\omega_0^2\left[\left(r-\dfrac{GM}{c^2}\right)^2-\left(\dfrac{GM}{c^2}\right)^2\right]\Psi,\notag
\\&\approx-\dfrac{\hbar^2}{2m}\nabla^2\Psi+\dfrac{1}{2}m\omega_0^2r^2\Psi-\dfrac{1}{2}m\omega_0^2\left(\dfrac{GM}{c^2}\right)^2\Psi.
\end{align}

Since the quantum system is far from the source, it is possible to do the previous approximation. For the perturbed Hamiltonian $\hat{H}_p$ we have
\begin{equation}
\hat{H}_p=-\dfrac{GMm}{r}+\dfrac{4GME^{(0)}_n}{rc^2}.
\end{equation}

The principal Hamiltonian is that of an isotropic harmonic oscillator with spherical symmetry whose eigenfunctions are well known from QM. These solutions are obtained after applying the separation of variables method, just as in the previous case $\Psi_{nklj}(t,r,\theta,\phi)=$R$_{kl}(r)$Y$_{lj}(\theta,\phi)\Psi(t)$, where $\Psi(t)$ is the solution in eq.(\ref{time solution}). Here R$_{nl}(r)$ is the radial function for an isotropic harmonic oscillator R$_{kl}(r)=r^le^{-\gamma r^2}$L$^{(l+1/2)}_k(2\gamma r^2)$, with $\gamma=\dfrac{m\omega}{2\hbar}$ and L$^{p}_q(x)$ being the generalized (or associated) Laguerre polynomials. The set $(n,k,l,j)$ become the quantum numbers for the isotropic harmonic oscillator with spherical symmetry from QM.\newline

Thus, the first order correction for the energy  from eq.(\ref{energy correction})  is given by 
\begin{equation}
E_n=\hbar\omega_0\left(n+\dfrac{3}{2}\right)-\dfrac{1}{2}m\omega_0^2\left(\dfrac{GM}{c^2}\right)^2-GMm\left[1+\dfrac{4E^{(0)}_n}{mc^2}\right]\left\langle \dfrac{1}{r}\right\rangle,
\end{equation}
where $E^{(0)}_n=\hbar\omega_0\left(n+\dfrac{3}{2}\right)$ is a well-known result from QM. Also note that there is a degeneration of these numbers since $n=2k+l$. A more general solution of $\left\langle \dfrac{1}{r}\right\rangle$ was found in \cite{Amrmstrong}, that for the case we are discussing reduces to the expression
\begin{align}
\label{iho correction}
\left\langle \dfrac{1}{r}\right\rangle=&\sqrt{\dfrac{m\omega_0}{\hbar}}\Gamma(l+1)[\dfrac{1}{2}(n-l-1)]!\times
\\&\sum_t\dfrac{(-1)^t}{[1/2(n-l-1)-t]!\Gamma(1/2(2l+1)+t+1)}\binom{-1/2}{t}^2\notag,
\end{align}
where $\Gamma(z)$ is the Gamma function $\Gamma(z)=\int^\infty_0t^{z-1}\exp(-t)dt$. When only the first term is taken from the sum of eq.(\ref{iho correction}), we obtain the energy value
\begin{align}
\label{correction oscillator}
E_n=&\hbar\omega\left(n+\dfrac{3}{2}\right)-\dfrac{1}{2}m\omega_0^2\left(\dfrac{GM}{c^2}\right)^2\notag\\
&-GMm\sqrt{\dfrac{m\omega_0}{\hbar}}\left[1+\dfrac{4E^{(0)}_n}{mc^2}\right]\Gamma(l+1)[\dfrac{1}{2}(n-l-1)]!.
\end{align}

Note that if $G=0$, we return to the solution for an isotropic harmonic oscillator with spherical symmetry without gravitational field. It is peculiar that the Schwarzschild radius $r_{\mathrm{sch}}$ is obtained in a very natural way.\newline

We are going to take typical experimental values for the case of a harmonic oscillator \cite{oscillator}. If we use a frequency of the order of $\omega_0\sim G$ Hz,the order of the mass $m = m_e$ that is the electron mass, and a total mass of the gravitational body $M\sim10^{16}$ kg, we obtain that the dominant term of the first order energy correction is $\sim10^{-5}E_{\mathrm{osc}}^{(0)}$ in the equation (\ref{correction oscillator}), while other contributions are of much lower order $<10^{-9}E^{(0)}_{\mathrm{osc}}$. We can conclude that an experiment for a harmonic oscillator with these values could be feasible in a laboratory.\newline

The two cases we have analyzed in this section are extreme, and correspond to those where the gravitational field is much more intense than the harmonic oscillator and vice-versa. The case where the two potentials are comparable is much more complex and we leave it for a future work.\newline

\section{Infinite Spherical Well}
\label{Infinite Spherical Potential G}
\qquad In this section we continue discussing well-known examples from QM with the novelty of the presence of a gravitational field. We now deal with the problem of an infinite spherical well barrier $V_{\mathrm{inf}}(r)$ that has two regions, this problem is commonly used in experimental verification or applications of QM. The barrier potential is given by 
\begin{equation}
V_{\mathrm{inf}}(r)= \left\{
	       \begin{array}{ll}
		0\ \ \ \text{if} \ \ \ 0<r<a, \\
		 \infty \ \ \ \text{otherwise}. \\
	       \end{array}
	     \right.
\end{equation}

We are only interested in studying the region where $0<r<a$, since outside of this region the wave function is zero, thus the probability to find a particle in the $r>a$ is null. The equation of motion (\ref{KG-N motion equation}) in the region of interest is
\begin{equation}
-\dfrac{\hbar^2}{2m}\nabla^2\Psi+mU\Psi -\dfrac{2\hbar^2U}{mc^2}\nabla^2\Psi=E\Psi
\end{equation}

We can choose a principal Hamiltonian operator $\hat{H}_0$ from the motion equation, where 
\begin{equation}
\hat{H}_0\Psi=-\dfrac{\hbar^2}{2m}\nabla^2\Psi=E^{(0)}\Psi.
\end{equation}

The zero-order correction of the energy is then $E^{(0)}_{ln}=\dfrac{\hbar^2}{2m}\dfrac{q_{ln}^2}{a^2}$. We can define the perturbed Hamiltonian $\hat{H}_p$ from the KG equation 
\begin{align}
\hat{H}_p=-\dfrac{GMm}{r}\left(1+\dfrac{4E^{(0)}}{mc^2}\right).
\end{align}

The first-order correction of the energy $E^{(1)}$ can be calculated using the zero-order correction of the eigenfunction $\Psi_{nlk}^{(0)}(r,\theta,\phi)=$A$_{ln}$j$_l\left(\dfrac{\mathrm{q}_{ln}}{a}r\right)$Y$_{lj}(\theta,\phi)$. Here q$_{ln}$ is n-th root of the Bessel spherical functions j$_l(x)$ and A$_{ln}$ is the constant of normalization 
\begin{equation}
\mathrm{A}_{ln}^2=\dfrac{2}{a^3[\mathrm{j}_{l+1}(\mathrm{q}_{ln})]^2}.
\end{equation}

Therefore,
\begin{equation}
\label{first order infinity sph energy}
E_{ln}^{(1)}=-GMm\left(1+\dfrac{4E^{(0)}}{mc^2}\right)\left\langle \dfrac{1}{r}\right\rangle,
\end{equation}
where, from perturbation theory we have 
\begin{equation}
\left\langle \dfrac{1}{r}\right\rangle=\mathrm{A}_{ln}^2\int_0^{a}\abs{\mathrm{j}_l\left(\dfrac{\mathrm{q}_{ln}}{a}r\right)}^2r\mathrm{d}r.
\end{equation}

Unfortunately, it is not possible to solve the last integral with analytic methods, thus we will integrate it numerically. In general, there exists for each state $n=2l+1$ degeneration, thus for $l=0$, we have that j$_0(x)=\sin(x)/x$, whose n-th root is q$_{0n}=n\pi$ and the normalization's constant is A$_{01}=\sqrt{2/a}$. Hence
\begin{equation}
\left\langle \dfrac{1}{r}\right\rangle\approx\dfrac{2a}{\pi^2}(1.218).
\end{equation}

Now, for $l=1$ the first excited state is three-fold degenerate, which means we need the first three roots of j$_1(x)=\dfrac{\sin(x)}{x^2}-\dfrac{\cos(x)}{x}$. The roots are q$_{11}\approx 4.49340$, q$_{12}\approx 7.72525$ and q$_{13}\approx 10.90412$.\newline 
Therefore for $n=1$
\begin{equation}
\left\langle \dfrac{1}{r}\right\rangle\approx \mathrm{A}_{11}^2\left( \dfrac{a}{\mathrm{q}_{11}}\right)^2 (0.4124),
\end{equation}
for $n=2$
\begin{equation}
\left\langle \dfrac{1}{r}\right\rangle\approx \mathrm{A}_{12}^2\left( \dfrac{a}{\mathrm{q}_{12}}\right)^2 (0.4590),
\end{equation}
and for $n=3$
\begin{equation}
\left\langle \dfrac{1}{r}\right\rangle\approx \mathrm{A}_{13}^2\left( \dfrac{a}{\mathrm{q}_{13}}\right)^2 (0.4778).
\end{equation}

We can continue this process for the next excited states, as in the previous cases. If we make $G=0$, we return to solutions for QM without gravitational field.\newline

Using the typical experimental values\cite{barrier} for a particle with an electron mass confined in an infinite well barrier of width $a\sim 10$nm, it is possible to obtain that $E_{\mathrm{inf}}^{(0)}\sim 10$eV. The dominant term in Eq.(\ref{first order infinity sph energy}) is $\sim 10^{-5}E_{\mathrm{inf}}^{(0)}$ for the first-order correction of energy, and the other term in this correction is $ <10^{-20}E_{\mathrm{inf}}^{(0)}$.

\section{Spherical Potential Barrier}

\qquad The problem of square well potential with certain symmetry (spherical, cartesian or cylindrical) is important for some experiments of quantum systems. In this section, we study a square well barrier with spherical symmetry. Similarly as in the previous sections, we analyze the case of QM for a square well potential with space-time curvature. We use the Newtonian metric from eq.(\ref{Newtonian metric}), so the equation of motion eq.(\ref{KG-N motion equation}) transforms into   
\begin{equation}
\label{spherical KG motion eq}
-\dfrac{\hbar^2}{2m}\nabla^2\Psi+mU\Psi +V_B\left(1+\dfrac{2U}{c^2}\right)\Psi-\dfrac{4U}{c^2}\dfrac{\hbar^2}{2m}\nabla^2\Psi=E\Psi,
\end{equation}
where again $U=-GM/r$ is the gravitational potential. We take the Laplacian operator $\nabla^2$ in spherical coordinates and the potential $V_\mathrm{B}$ as 
\begin{equation}
V_\mathrm{B}(r)= \left\{
	       \begin{array}{ll}
		-U_0\ \ \ \text{if} \ \ \ r<a, \\
		 0 \ \ \ \ \ \ \ \text{if} \ \ \ r>a,
	       \end{array}
	     \right.
\end{equation}

Thus, from Eq.(\ref{spherical KG motion eq}) we can identify a principal Hamiltonian operator $\hat{H}_0$ 
\begin{equation}
\hat{H}_0\Psi=-\dfrac{\hbar^2}{2m}\nabla^2\Psi+V_\mathrm{B}\Psi=E^{(0)}\Psi.
\label{perturbed sph barrier}
\end{equation}

The perturbed Hamiltonian $\hat{H}_p$ can be defined from eq.(\ref{spherical KG motion eq})   
\begin{equation}
\hat{H}_p=-\dfrac{GMm}{r}\left(1+\dfrac{4E^{(0)}}{mc^2}-\dfrac{2V_\mathrm{B}}{mc^2}\right)
\end{equation}

In general, we can find the fist-order correction of energy from eq.(\ref{energy correction}) with the perturbed Hamiltonian operator in eq.(\ref{perturbed sph barrier}). This yields
\begin{equation}
E^{(1)}=-GMm\left(1+\dfrac{4E^{(0)}}{mc^2}-\dfrac{2V_\mathrm{B}}{mc^2}\right)\left\langle \dfrac{1}{r}\right\rangle.
\end{equation}

The case where $E<0$ shows the quantum nature of the system due to the fact that the energy spectrum is discrete. Potential $V_B$ defines naturally two regions, $Region\ I$ ($r<a$) and $Region\ II$ ($r>a$). Both regions without gravitational field are well-known. For $Region\ I$ the first-order correction of energy $E^{(1)}_{I}$ is given by
\begin{equation}
E^{(1)}_I=-GMm\left(1+\dfrac{4E_{I}^{(0)}}{mc^2}+\dfrac{2U_0}{mc^2}\right)\left\langle \dfrac{1}{r}\right\rangle_I,
\end{equation}
where the zero-order correction of the eigenvalue is $E_{I}^{(0)}=\dfrac{\hbar^2}{2m}(\gamma_{ln})^2$ and the zero-order correction eigenfunction reads $\Psi_{lnj}(r,\theta,\phi)=$R$_{ln}(r)$Y$_{lp}(\theta,\phi)$. As usual, Y$_{lp}(\theta,\phi)$ are the spherical harmonic functions and R$_{ln}(r)=$A$_{ln}j_l(\gamma_{ln}r)$ is the radial solution, such that j$_l(x)$ are the spherical Bessel functions. Also k$_1=\gamma_{ln}$ is the n-th solution of the transcendental equation due to the boundary and continuity condition given by
\begin{equation}
\label{trascendental eq}
\dfrac{1}{\mathrm{h}_l^{(1)}(i\mathrm{k}_2r)}\dfrac{\mathrm{d}\mathrm{h}_l^{(1)}(i\mathrm{k}_2r)}{\mathrm{d}r}|_{r=a}=\dfrac{1}{\mathrm{j}_l(ik_1r)}\dfrac{\mathrm{d}\mathrm{j}_l(i\mathrm{k}_1r)}{\mathrm{d}r}|_{r=a}.
\end{equation}

For $l=0$, it reduces to the transcendental equation
\begin{equation}
-\mathrm{k}_2=\mathrm{k}_1\cot(\mathrm{k}_1a).
\end{equation}

In the limit $|E|<<U_0$, we return to the solution for the first root. When $l=0$ we recover the same result as in the last section, namely k$_1a\approx \pi/2$. In general for Region I, we need to integrate
\begin{equation}
\left\langle \dfrac{1}{r}\right\rangle_I=\mathrm{A}_{ln}^2\int^a_0\abs{\mathrm{j}_l\left(\dfrac{\sigma_{ln}}{a}r\right)}^2r\mathrm{d}r,
\end{equation}
where $\sigma_{ln}=a\gamma_{ln}$. If we define the parameter $\eta=\sigma_{ln}/$q$_{ln}$ that compares the n-th solution of the transcendental equation with the n-th root, we see that when $\eta=1$ we return to case of Section \ref{Infinite Spherical Potential G}. However, the solution of this integral should be calculated using numerical methods.\newline

In $Region\ II$, the zero-order correction is given by
\begin{equation}
E_{II}^{(0)}=\dfrac{\hbar^2}{2m}(\gamma^*_{ln})^2=\dfrac{\hbar^2}{2M}\left(\dfrac{\sigma^*_{ln}}{a}\right)^2,
\end{equation}
where we have introduced the new parameters $\sigma^*_{ln}=a\gamma^*_{ln}$, and $\eta^*=\gamma^*_{ln}/$q$_{ln}$.\newline

The first-order correction is similar to that of $Region\ I$
\begin{equation}
E^{(1)}_{II}=-GMm\left(1+\dfrac{4E_{II}^{(0)}}{mc^2}\right)\left\langle \dfrac{1}{r}\right\rangle_{II},
\end{equation}
with the eigenfunction $\Psi_{lnj}(r,\theta,\phi)=$R$_{ln}(r)$Y$_{lj}(\theta,\phi)$. Nevertheless, the radial function has a new form R$_{ln}(r)=B$h$_l^{(1)}(i\gamma_{ln}^*r)=B$j$_l(i\gamma_{ln}^*r)+i B$n$_l(i\gamma_{ln}^*r)$, where $B$ is a constant of normalization, h$_l^{(1)}(x)$ are the spherical Hankel functions of the first kind, n$_l(x)$ are the spherical Neumann function and $k_2=\gamma_{ln}^*$ is the solution of the transcendental equation (\ref{trascendental eq}). Therefore
\begin{align}
\left\langle \dfrac{1}{r}\right\rangle_{II}=&B^2\int_a^\infty\abs{\mathrm{h}_l^{(1)}\left(i\gamma^*_{ln}\right)}^2r\mathrm{d}r\notag\\
=&B^2\int_a^\infty\left(\abs{\mathrm{j}_l(i\gamma_{ln}^*r)}^2- \abs{\mathrm{n}_l(i\gamma_{ln}^*r)}^2\right) r\mathrm{d}r.
\end{align}

We observe that outside the finite potential barrier the wave function attenuates as it occurs in standard QM.\newline 

The solution of this perturbation is the most general one, and if we want to solve the previously integral, we need to set every parameter for a specific case. As expected, just like in the previous cases, if $G=0$ we return to the solutions without a gravitational field.

\section{Finite Spherical Well}
\qquad Other important and fundamental well-known problem in standard QM is to analyze a finite barrier potential. In our case we study the problem with spherical symmetry, instead of the typical one-dimensional solutions\cite{Cohen:2005}\cite{Cohenv2:2005}\cite{Zettilli:2009}\cite{Pena:2006}. The principal issue in this problem is the analysis of the tunneling effect, which is produced because the potential $V_T(r)$ has three regions. In the regions where there is not a finite potential could be observed the tunneling effect, which is the transmission of the wave function between these regions. The spherical finite potential $V_T(r)$ can be read as
\begin{equation}
\label{tunneling potential}
V_\mathrm{T}(r)= \left\{
	       \begin{array}{lll}
		0 \ \ \ \text{if} \ \ \ r<a, \\
		U_0\ \ \text{if}\ \ \ a<r<2a,\\
		 0 \ \ \ \  \text{if} \ \ \ r>2a,
	       \end{array}
	     \right.
\end{equation}
where $U_0>0$ and $a>0$, we can divide in three regions where the potential affects. The first region for $0<r<a$, the second region for $a<r<2a$, and the third region is for $2a<r$. We again use the Newtonian metric from eq.(\ref{Newtonian metric}), and we transform eq.(\ref{KG-N motion equation}) with the potential (\ref{tunneling potential}). Thus, the non-perturbed Hamiltonian $H_0$ is given by 
\begin{equation}
\label{H0 finite barrier}
\hat{H}_0\Psi=-\dfrac{\hbar^2}{2m}\nabla^2\Psi+V_\mathrm{T}\Psi=E^{(0)}\Psi.
\end{equation}
Furthermore, we can identify the perturbed Hamiltonian as
\begin{equation}
\label{finite perturbed hamiltonian}
\hat{H}_p=-\dfrac{GMm}{r}\left(1+\dfrac{4E^{(0)}}{mc^2}-\dfrac{2V_\mathrm{T}}{mc^2}\right),
\end{equation}
We can calculate the first order correction, as we just saw before
\begin{equation}
E^{(1)}=-GMm\left(1+\dfrac{4E^{(0)}}{mc^2}-\dfrac{2V_\mathrm{T}}{mc^2}\right)\left\langle \dfrac{1}{r}\right\rangle.
\end{equation}

Thus, we need to solve three differential equations with the corresponding boundary conditions for each case of the eigenvalues equation with the non-perturbed Hamiltonian which are given by 
\begin{align}
    \nabla^2\Psi_I^{(0)}&=-k_I^2\Psi_I^{(0)},\\
    \nabla^2\Psi_{II}^{(0)}&=-k_{II}^2\Psi_{II}^{(0)},\\
    \nabla^2\Psi_{III}^{(0)}&=-k_{III}^2\Psi_{III}^{(0)},
\end{align}
where $I$, $II$, $III$ denoted each region, the Laplacian has a spherical symmetry. Moreover, $k^2_I=2mE^{(0)}_I/\hbar^2$, $k^2_{II}=2m(U_0-E^{(0)}_{II})/\hbar^2$ and $k^2_{III}=2mE^{(0)}_{III}/\hbar^2$. We can write the solution to zero order of the wave function as $\Psi^{(0)}(r,\theta,\phi)=$R$_{pl}^{(0)}(r)$Y$_{lj}(\theta,\phi)$, where Y$_{lj}$ are the harmonic spherical functions, and R$_{pl}$ is the radial function for each region. The boundary and continuity conditions are given by
\begin{align}
    \mathrm{R}_I^{(0)}(r=a)&=\mathrm{R}_{II}^{(0)}(r=a),\\
    \mathrm{R}_{II}^{(0)}(r=2a)&=\mathrm{R}_{III}^{(0)}(r=2a),\\
    \dfrac{1}{\mathrm{R}^{(0)}_I}\dfrac{\mathrm{d}\mathrm{R}^{(0)}_I}{\mathrm{d}r}|_{r=a}&=\dfrac{1}{\mathrm{R}^{(0)}_{II}}\dfrac{\mathrm{d}\mathrm{R}^{(0)}_{II}}{\mathrm{d}r}|_{r=a},\\
    \dfrac{1}{\mathrm{R}^{(0)}_{II}}\dfrac{\mathrm{d}\mathrm{R}^{(0)}_{II}}{\mathrm{d}r}|_{r=2a}&=\dfrac{1}{\mathrm{R}^{(0)}_{III}}\dfrac{\mathrm{d}\mathrm{R}^{(0)}_{III}}{\mathrm{d}r}|_{r=2a},
    \label{boundary condition tunneling}
\end{align}
where R denoted the radial function in each region. With the boundary and continuity conditions we can solved the radial solutions as follows
\begin{align}
    \mathrm{R}^{(0)}_I&=\mathrm{A}_{pl}\mathrm{j}_l(\alpha_{pl}r)+\mathrm{B}_{pl}\mathrm{n}_l(\alpha_{pl}r),\\
    \mathrm{R}^{(0)}_{II}&=\mathrm{C}_{pl}\mathrm{j}_l(\beta_{pl}r)+\mathrm{D}_{pl}\mathrm{n}_l(\beta_{pl}r),\\
    \mathrm{R}^{(0)}_{III}&=\mathrm{E}_{pl}\mathrm{j}_l(\gamma_{pl}r)+\mathrm{F}_{pl}\mathrm{n}_l(\gamma_{pl}r),
\end{align}
where A$_{pl}$, B$_{pl}$, C$_{pl}$, D$_{pl}$, E$_{pl}$, F$_{pl}$, and G$_{pl}$ are complex scalars which are used to normalize the wave functions. Furthermore, $\alpha_{pl}$, $\beta_{pl}$, and $\gamma_{pl}$ are the solutions to boundary and continuity conditions which are above. Moreover, j$_{l}$(x) and n$_{l}$(x) are the spherical Bessel functions.\newline

In addition, we can calculate the first order correction to the wave function where we use the perturbation method, such correction is given by

\begin{equation}
    \label{first order wave function}
   \ket{\psi^{(1)}_i}=\ket{\psi^{(0)}_i}+\sum_{k\neq i}\dfrac{\bra{\psi^{(0)}_k}\hat{H}_p\ket{\psi^{(0)}_i}}{E_i^{(0)}-E_k^{(0)}}\ket{\psi^{(0)}_k}
\end{equation}
here, the eigenvalues to zero-order are given for each region by $E^{(0)}_I=\dfrac{\hbar}{2m}\alpha_{pl}^2$,  $E^{(0)}_{II}=\dfrac{\hbar}{2m}\beta_{pl}^2$ and  $E^{(0)}_{III}=\dfrac{\hbar}{2m}\gamma_{pl}^2$. In fact, we note the solutions are numerical, because the boundary conditions are transcendental equations. We can reduce the problem to solve that equations and to give an expression for each region to $\bra{j}r^{-1} \ket{k}$ to zero-order eigenvectors.\newline

As we know, the tunneling effect is, due to there exists a probability to find the wave in the third region when $E^{(0)}<U_0$. The tunneling probability or transmission probability $T$ is the ratio of the transmitted intensity to the incident intensity. If we take into account the correction to first order into the wave function, we can calculate the tunneling probability as follows
\begin{equation}
    T=\dfrac{|\Psi_{tra}|^2}{|\Psi_{in}|^2}=\dfrac{|\Psi^{(0)}_{tra}+\Psi^{(1)}_{tra}|^2}{|\Psi^{(0)}_{in}+\Psi^{(1)}_{in}|^2},
\end{equation}
the transmitted wave is related with the wave function at region III, and the incident wave is related with the region I. Thus, the correction in both cases is of the same order. Furthermore, we need to find the tunneling probability which is the ratio to quadratic first order eigenfunctions. Hence, the contribution to the probability correction is smaller than in the other problems showed in this paper. The energy values could be measured at the same order as the other examples that we studied here.  

\section{Conclusions}

\qquad Throughout this work, we studied the KG equation in a weak gravitational field for different external potentials to understand the quantum effects of a boson gas on a gravitational field. Typical examples of QM were analyzed featuring the addition of space-time curvature (or gravitational effects). Starting from the most general equation for bosons in Quantum Field Theory in curved space-times, we found a generalized Schr\"odinger equation that is simply the KG covariant one. To solve the differential equation, we identified, for the spatial part, the principal and perturbed Hamiltonian operator in each case, and compared them with the well-known results in QM. For the time-dependence part, we solved the differential equation taking account that in the classical limit we needed to return to the solution in QM for the same case. Each example was worked out on an inertial frame. This is an important aspect to highlight due to the main idea of this work. We expect to obtain the same results when gravitational effects are measured on a quantum system in the laboratory. This quantum system (with scalar particles) will be on a non-inertial frame, where we predict that the Einstein equivalence principle gives us a correspondence between experimental and theoretical results.\\

These results let us find the limit, where we could measure gravitational effects on a boson system, namely $2\alpha\, mM/m_{pl}^2\sim 1$. For this analysis, we could be able to apply our results in micro black holes scale\cite{microBH}. For example, if we consider a mass of a scalar particle $m$ as the mass of the electron, we obtain that the gravitational mass $M$ that affects the electron by quantum gravitational effects is $\sim 10^{16}$kg. In the same way, a $M\sim 10^{12}$kg could affect an alpha particle. We want to relate the theoretical and experimental results. This connection can be done by means of the EEP comparing the repercussions of a boson system into an inertial frame with gravitational interaction and into a non-inertial frame without gravitational field.\newline 

As a consequence, if we think on the easiest non-inertial frame, we could think on a rotating system, for which there would not be a real mass $M$, but rather an effective mass given by angular frequency $\Omega$, where $\Omega^2=GM/r^3$ into a Newtonian approximation. Using a detector to $r=1$ meter, the corresponding gravity can be reached with a rotational wheel spinning with an angular frequency of $\Omega\approx 1543$ rad/s=245 rev/s for $M\sim 10^{16}$kg for electron mass, and $\Omega\approx 19$ rad/s=3 rev/s for $M\sim 10^{12}$kg for the case of alpha particles. However, for a mass $M\sim M_\odot$, we would need an angular frequency of $\Omega\sim 10^{9}$rev/s, to obtain relativistic effects.\newline 

Although, as a dimensional approximation, the previous analysis can be acceptable. Nevertheless, we need to pay attention to other effects, such as the Coriolis force and the centrifugal force (see for more detail on\cite{RRFrames:2004}) where we take a rotating frame $F'$ with the angular velocity $-\mathbf{\Omega}$ relative to the inertial frame $F_0$. We can obtain the Hamiltonian
\begin{equation}
    \label{hmailtonin Coriolis}
    H=\dfrac{1}{2m}(\mathbf{\hat{p}}-m\mathbf{\Omega}\times\mathbf{x})^2-\dfrac{1}{2}m(\mathbf{\Omega}\times\mathbf{x})^2,
\end{equation}
where the velocity $\mathbf{V}$ is, now, given by $\mathbf{\Omega}\times\mathbf{x}$.\newline

Analyzing the typical experiments on a harmonic oscillator \cite{oscillator} and on an infinite well barrier \cite{pozo} for a particle with an electron mass, the correction due to the gravitational effects has, in both cases, a dominant term which is given by $GMm\left\langle \dfrac{1}{r}\right\rangle$. Taking the typically values of an experimental result for the case of a harmonic oscillator with frequency $\omega_0\sim G$Hz and using the value for the mass $M\sim 10^{16}$kg with the dominant term of the first-order correction of the energy, in this case is $\sim 10^{-5}E_{\mathrm{osc}}^{(0)}$ from Eq.(\ref{correction oscillator}), while other contributions in this equation are $<10^{-9}E^{(0)}_{\mathrm{osc}}$, we can conclude an experiment for harmonic oscillator with these values could be feasible in a lab. Furthermore, for a particle with an electron mass confined in an infinite well barrier of width $a\sim 10$nm, it is possible to obtain that $E_{\mathrm{inf}}^{(0)}\sim 10$eV. The dominant term in Eq.(\ref{first order infinity sph energy}) is $\sim 10^{-5}E_{\mathrm{inf}}^{(0)}$ for the first-order correction of energy, and the other term in this correction is $ <10^{-20}E_{\mathrm{inf}}^{(0)}$. With the technology today is possible to realize these measurements.\newline 

Note that in either cases, the first-order corrections of energy have the same order if we compare them with their zero-order counterparts. With this analysis, it is possible to conclude that we can measure in a laboratory the effects of the quantization of a weak gravitational field directly or using non-inertial systems.\newline

We thought of a way to do experiments to verify the results presented in this work. On the one hand, a rotating laboratory can be used, where the data is somehow sent abroad. In the rotating laboratory we do the typical quantum exercises described here, the rotation of the laboratory creates artificial gravity that we can modify by varying the rotation frequency, using the approximations found in the previous discussion. On the other hand, we must compare these with the results made by the same quantum experiments, but which will be carried out in a research laboratory in Earth orbit \cite{ObservationofBoseEinstein:2020nature}, where gravity is zero. We can then compare these three cases where non-inertial, inertial and weak gravity frames of reference have been used. We hope to deliver results in this direction one day.\newline

\section{Acknowledgments}
\qquad This work was partially supported by CONACyT M\'exico under grants CB-2011 No. 166212, CB-2014-01 No. 240512, Project
No. 304001, 269652 and Fronteras Project 281; The authors are gratefully for the computing time granted by LANCAD and CONACYT in the Supercomputer Hybrid Cluster "Xiuhcoatl" at GENERAL COORDINATION OF INFORMATION AND COMMUNICATIONS TECHNOLOGIES (CGSTIC) of CINVESTAV. URL: https://clusterhibrido.cinvestav.mx/ and Abacus clusters at Cinvestav, IPN; I0101/131/07 C-234/07 of the Instituto Avanzado de Cosmolog\'ia (IAC) collaboration (http:// www.iac.edu.mx). O.G. acknowledges financial support from CONACyT doctoral fellowship and appreaciates Angelica C. Aguirre Castañ\'on for her valuable review and support. Works of T.M. are partially supported by Conacyt through the Fondo Sectorial de Investigaci\'on para la Educaci\'on, grant CB-2014-1, No. 240512.
\newline

The data that support the findings of this study are available from the corresponding author upon reasonable request.
\bibliographystyle{ieeetr}
\bibliography{ref.bib}
\end{document}